\documentstyle[multicol,aps,prl,graphicx]{revtex}

\begin{document}

\newcommand{\bc}{\begin{center}}
\newcommand{\ec}{\end{center}}
\newcommand{\be}{\begin{equation}}
\newcommand{\ee}{\end{equation}}
\newcommand{\beqn}{\begin{eqnarray}}
\newcommand{\eeqn}{\end{eqnarray}}

\begin{multicols}{2}
\narrowtext

\parskip=0cm

\noindent
{\large\bf Comment on "Disorder Induced Quantum Phase Transition in
  Random-Exchange Spin-1/2 Chains"} 
\medskip

In a recent Letter Hamacher, Stolze and Wenzel \cite{prl} studied
the disordered spin-1/2 AFM XXZ chain 
\be
H=J\sum_{i=1}^{L-1} \Bigl[\lambda_i(S_i^xS_{i+1}^x+S_i^yS_{i+1}^y)+
\Delta S_i^zS_{i+1}^z\Bigr]
\ee
with $\lambda_i$ i.i.d. random variables uniformly distributed over
the interval $[1-W;1+W]$, the parameter $W$ controlling the strength
of the disorder. The authors claim that for $\Delta<1$ they found
numerical evidence for non-universal behavior for weak disorder ($W<1$)
manifested in a continously varying exponent $\eta(W)$ describing the
asymptotic decay of the transverse spin correlations 
\be
C^{xx}(r)=\langle S_i^x S_{i+r}^x \rangle \propto r^{-\eta(W)}\;,
\ee
where $\langle\ldots\rangle$ denotes the ground state expectation
value averaged over the disorder and the sites $i$. They concluded
that there is no universal infinite randomness fixed point (IRFP) as
predicted by D.\ Fisher \cite{fisher}.

In this comment we show that these conclusions are wrong and that the
numerical data presented in \cite{prl} are not in the asymptotic
regime. This misconception is due to the fact that the authors ignored
the existence of a $W$-dependent crossover length scale $\xi_W$ that
describes the crossover from the pure fixed point to the only relevant
IRFP: For $L\ll\xi_W$ one observes the critical behavior of the pure
system ($\lambda_i=const.$), and only for $L\gg\xi_W$ the true
asymptotic critical behavior ($\eta(W)=2$, independent of $W>0$)
of the disordered chain becomes visible. Even for strong disorder
($W=1.0$) $\xi_W$ is of the same order of magnitude as the system
sizes considered in \cite{prl}.

In order to be able to reach sufficiently large system sizes we
restrict ourselves to $\Delta=0$ in which case (1) reduces to a free
fermion model and the ground state computations are done following
\cite{fermion}. In Fig.\ 1 we show the averaged bulk correlation
function $C^{xx}(L/2)$ for different strengths of the disorder. We
observe that asymptotically (i.e.\ for $L\to\infty$) the data follow
the behavior $C^{xx}(L)\propto L^{-2}$ as predicted by the real space
renormalization group \cite{fisher}. Only for small $L$ the data {\it
appear} to follow a non-universal (i.e.\ $W$-dependent) power law, and
this is the region on which \cite{prl} reports.

Due to the presence of a crossover length scale $\xi(W)$ the
correlation function obeys the scaling form
\be
C^{xx}(L/2)=L^{-1/2}\tilde{c}(L/\xi_W)
\ee
where $\tilde{c}(x){\rm const}$ for $x\to0$, and $\tilde{c}(x)\to
x^{-3/2}$ for $x\to\infty$. This implies $C^{xx}(L/2)\propto L^{-1/2}$
for $L\ll\xi_W$ (the pure behavior) and $C^{xx}(L/2)\propto L^{-2}$ for
$L\gg\xi_W$ (the IRFP behavior). In the inset of Fig.\ 1 we show such
a scaling plot of the data in the main figure. We have chosen $\xi_W$
for $W=1$ such that the crossover region is centered around
$L/\xi_W\approx1$, the other estimates for $\xi_W$ are then chosen to
give the best data collapse. We see that for all disorder strength the
maximum system sizes used in \cite{prl} are still well within the
crossover region and {\it not} in the asymptotic regime. We do not
expect that this situation will improve for $\Delta>0$. In general
$\xi_W$ diverges when $W\to0$, and we found that our estimates for
$\xi_W$ obey $\xi_W\propto\delta_W^{-\Phi}$ where $\delta$ is the
($W$-dependent) variance of the random variable $\ln\lambda_i$, which
determines according to \cite{fisher} the RG-flow of the model (1). We
obtain $\Phi\approx1.24\pm0.02$.

\begin{figure}[t]
\includegraphics[width=\columnwidth]{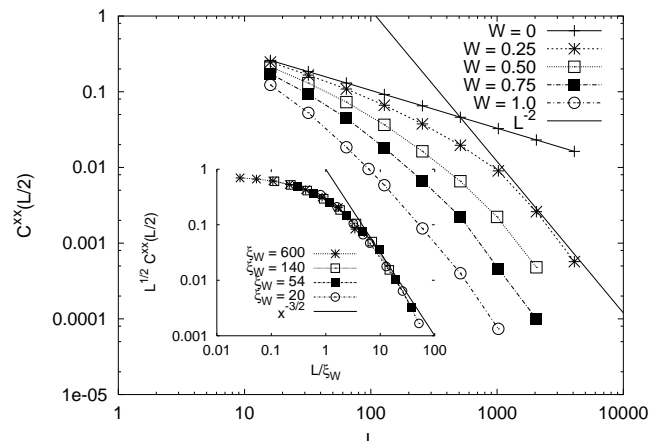}
\caption{Averaged correlation function $C^{xx}(L/2)$ as a function of
  the system size $L$ in a log-log-plot for W=0, 0.25, 0.5, 0.75, 1.0
  (top to bottom). The data are averaged over 50000 for
  $L\le1024$, 3000 for L=2048 and 500 for L=4096, the
  statistical error is smaller than the symbol sizes. The data for the
  pure system ($W=0$) follow $C^{xx}(L/2)\propto L^{-1/2}$, the full
  line with slope $-2$ is the expected asymptotic behavior according
  to the IRFP scenario [2]. {\bf Inset:} Scaling plot according to
  Eq. (3) for the data of the main figure with $\xi_W=600$, $140$,
  $54$, $20$ for $W=0.25$, $0.5$, $0.75$ and $1.0$, respectively}
\end{figure}

To conclude we have shown that contrary to what has claimed in
\cite{prl} there is no non-universal behavior in model (1) for weak
disorder and that by taking into account the existence of a crossover
length scale the numerical data, ours as well as theirs, for
$\Delta<1$ are compatible with the IRFP scenario predicted in
\cite{fisher}.
\medskip

\noindent
Nicolas Laflorencie$^a$ and Heiko Rieger$^b$\\
{\small
\begin{tabular}{cl}
$^a$ & Laboratoire de Physique Quantique;\\
     & IRSAMC; Universit\'e Paul Sabatier;\\
     & 31062 Toulouse; France;\\
$^b$ & Theoretische Physik; Universit\"at des Saarlandes;\\
     & 66041 Saarbr\"ucken; Germany
\end{tabular}
}
\medskip

\noindent
Date: 2 December 2002

\end{multicols}

\end{document}